\documentclass[aps,pra,reprint,superscriptaddress]{revtex4-2}

\usepackage{graphicx}
\usepackage{bm}
\usepackage{amsmath,amsthm,amssymb,amscd}
\usepackage[colorlinks,linkcolor=blue, urlcolor=blue, anchorcolor=blue, citecolor=blue]{hyperref}
\DeclareMathAlphabet{\altmathcal}{OMS}{cmsy}{m}{n}
\usepackage{microtype}
\usepackage{xcolor}
\newcommand{\cor}[1]{{\color{black} #1}}

\usepackage[normalem]{ulem}

\begin{document}


\title{Watt-level coherent microwave emission from dissipation engineered solid-state quantum batteries}


\author{Yuanjin Wang}
\affiliation{Center for Quantum Technology Research and Key Laboratory of Advanced Optoelectronic Quantum Architecture and Measurements (MOE), School of Physics, Beijing Institute of Technology, Beijing 100081, China}

\author{Hao Wu}
\email[]{hao.wu@bit.edu.cn}
\affiliation{Center for Quantum Technology Research and Key Laboratory of Advanced Optoelectronic Quantum Architecture and Measurements (MOE), School of Physics, Beijing Institute of Technology, Beijing 100081, China}

\author{Mark Oxborrow}
\affiliation{Department of Materials and London Centre for Nanotechnology, Imperial College London, SW7 2AZ London, United Kingdom}

\author{Qing Zhao}
\email[]{qzhaoyuping@bit.edu.cn}
\affiliation{Center for Quantum Technology Research and Key Laboratory of Advanced Optoelectronic Quantum Architecture and Measurements (MOE), School of Physics, Beijing Institute of Technology, Beijing 100081, China}


\date{\today}

\begin{abstract}
Recently proposed metastability-induced quantum batteries have shown particular promise for coherent microwave generation. However, achieving high-power coherent microwave generation in quantum batteries remains fundamentally challenging due to quantum correlations, aging, and self-discharging processes. For the cavity-quantum-electrodynamics (CQED)-based quantum batteries, a further trade-off arises between strong spin–photon coupling for energy storage and sufficient output coupling for power delivery. To overcome these constraints, we introduce dissipation engineering as a dynamic control strategy that temporally separates energy storage and release. By suppressing emission during charging and rapidly enhancing the output coupling during discharging, we realize nanosecond microwave bursts with watt-level peak power. By optimizing three dissipation schemes, we improve work extraction efficiency of the quantum battery by over two orders of magnitude and achieve high power compression factors outperforming the state-of-the-art techniques, establishing dissipation engineering as a pathway toward room-temperature, high-power coherent microwave sources.

\end{abstract}


\maketitle




\section{INTRODUCTION}
Quantum batteries, using quantum systems for efficient storage and controlled release of energy, represent a straightforward application scenario of quantum thermodynamics\cite{PhysRevE.87.042123}. Despite rapid theoretical progress\cite{RevModPhys.96.031001} and various proposed experimental platforms (e.g., nuclear/electronic spins\cite{PhysRevA.106.042601,PhysRevLett.133.180401,d9k1-75d4}, superconducting qubits\cite{Hu_2022}, ultracold atoms\cite{PhysRevA.110.032205} and organic excitons\cite{doi:10.1126/sciadv.abk3160,PhysRevE.111.044118}), demonstrations of useful functionalities of quantum batteries remain scarce. In particular, their integration into practical devices for energy conversion and power generation has only begun to be explored\cite{bhyh-53np,hymas2025experimentaldemonstrationscalableroomtemperature}. 


Coherent microwave generation is vital for quantum information processing, where various quantum systems, such as the superconducting circuits\cite{doi:10.1126/science.1231930}, neutral atoms\cite{Saffman_2016}, trapped ions\cite{PhysRevLett.113.220501}, and optically active defect centers in solids\cite{annurev-conmatphys-030212-184238,PhysRevLett.130.030601}, require high-standard microwave control and readout mechanisms\cite{9318753}. Recently, a metastability-induced quantum battery architecture was proposed as a promising route for coherent microwave generation by exploiting the masing process of the charged battery under ambient conditions\cite{73rk-6cp6}. The maser oscillator powered by the quantum battery not only bridges fundamental concepts of quantum thermodynamics and applications in quantum electronics, but also offers the intrinsic advantage of low-phase-noise microwave generation\cite{Belyaev:2023} due to the high coherence of the emitted photons and stable spin transitions in the battery.

In addition to coherence, the power scalability of the aforementioned quantum oscillator up to watt levels is also crucial for fast manipulation\cite{9318753,PhysRevA.110.053312,Froning2021,PhysRevX.13.031022} and high-fidelity readout\cite{10.1063/1.4922664,PhysRevX.14.041008,ohkuma2025coherentcontrolsolidstatedefect,Ramsay2023} of the microwave-based quantum systems. However, the realization of high-power coherent microwave generation faces fundamental challenges in the quantum battery system. The central difficulty lies in efficiently converting stored energy into coherent radiation, a process limited by correlations\cite{PhysRevLett.122.047702,10.1116/5.0184903}, aging\cite{PhysRevA.100.043833}, and self discharging\cite{PhysRevE.109.054132,PhysRevE.103.042118}. Moreover, for the oscillator that relies on the cavity quantum electrodynamics (CQED) mechanism, there is a trade-off between maintaining strong spin–photon coupling for efficient energy storage and enabling sufficient outcoupling for power delivery\cite{Breeze2017,Zollitsch2023}. If the cavity quality factor ($Q$) is too high, the stored inversion cannot be efficiently released; if it is too low, premature emission prevents the battery from accumulating enough energy for high-power operation. This tension, well known in laser and maser physics, also becomes critical for quantum batteries, where both energy storage and power extraction must be optimized simultaneously. 

Within this context, dissipation engineering\cite{PhysRevLett.77.4728,Marcos_2012} has emerged as a powerful tool for controlling energy flow in CQED systems by tailoring the system-environment coupling. This concept has recently gained significant attention in the field of nonlinear optics for the generation of non-Hermitian\cite{Pontula2025} and dark pulse\cite{Lv2025} combs, where the dissipation engineering was employed for introducing non-reciprocity or mitigating unnecessary nonlinear dynamics of energy flow.

In this work, we extend the concept of dissipation engineering from controlling nonlinear optical processes to modulating the discharge dynamics of quantum batteries. 
Specifically, we integrate a solid-state quantum battery with a microwave cavity to form a maser oscillator and design a controlled discharge system that efficiently releases stored microwave photons into the external environment. 
We investigate three different dissipation schemes for identifying the optimal modulation strategy in terms of the coherent microwave emission power. We show that with appropriate design of the modulation parameters including the modulation duration, minimum dissipation rate, and delay time, all three schemes are capable of generating nanosecond pulses with significantly increased output power up to watt levels. These results reveal the tunability of the discharge process of quantum batteries via dissipation engineering and lay the foundation for realizing high-power microwave coherent generation using dynamical CQED control approaches. 

\begin{figure}[tb!]
	\includegraphics[width=0.48\textwidth]{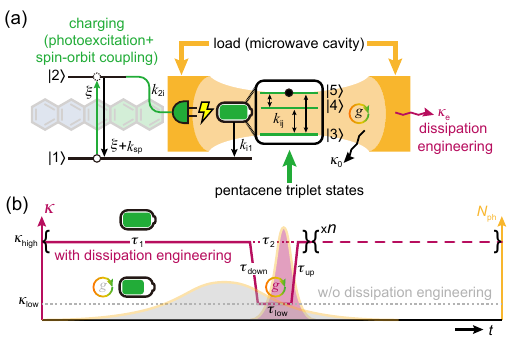}
	\caption{\label{fig1} (a) Schematic diagram of the charging and working mechanisms of the metastability-induced quantum battery. Dissipation engineering is realized by modulating the external coupling rate $\kappa_\textrm{e}$ of the microwave cavity. (b) Concept of using dissipation engineering for modulating the coherent microwave generation from the quantum battery. $N_\textrm{ph}$, microwave photon number.}
\end{figure}

\section{SYSTEM AND CONCEPT OF DISSIPATION ENGINEERING}

The working principle of the quantum battery-powered maser oscillator is shown in Fig.\hyperref[fig1]{1(a)}. The quantum battery is constituted by the metastable photoexcited triplet electron spins of the pentacene molecules doped in a crystalline \textit{p}-terphenyl matrix\cite{73rk-6cp6}. Photoexcitation of the pentacene molecules in the ground singlet state $|1\rangle$ to the excited singlet state $|2\rangle$, combined with the spin-orbit-coupling-induced intersystem crossing from $|2\rangle$ to the metastable triplet state can be considered as the charging process of the battery. 
In absence of the external magnetic field, the triplet spins possess three non-degenerate sublevels, denoted by $|3\rangle$, $|4\rangle$ and $|5\rangle$, due to the dipolar interactions. The sublevels $|3\rangle$ and $|5\rangle$ with a transition frequency $\omega_\textrm{35}$ around 1.45 GHz, characterized by the zero-field-splitting parameters\cite{10.1063/1.442520}, are considered as the states of the quantum battery for the energy storage and extraction. The work extraction manifests as the masing process when the charged quantum battery (i.e., the sublevel $|5\rangle$ is overpopulated) is coupled to a microwave cavity whose electromagnetic mode frequency $\omega_\textrm{m} =\omega_\textrm{35}$. The coupling can be quantified using the spin-photon coupling strength $g$. The complete process of the pentacene triplet mechanism for coherent microwave generation via masing can be found in the seminal work\cite{Oxborrow2012}. 

As an open quantum system, the battery can dissipate energy via the depopulation to $|1\rangle$ and spin-lattice relaxations within the triplet sublevels. The energy extracted from the battery to the load, i.e., the cavity, can also dissipate due to the intrinsic cavity loss and the external coupling for detecting the microwave photons generated in the cavity. The two dissipation channels with the respective rates $\kappa_{0}$ and $\kappa_\textrm{e}$ contribute to the total dissipation of the cavity, of which the rate can be characterized as $\kappa = \kappa_{0}+\kappa_\textrm{e}$. $\kappa_{0}$ is influenced by the ohmic and dielectric properties of the microwave cavity, whereas $\kappa_\textrm{e}$ is determined by the output coupling coefficient.

In this system, dissipation engineering is realized by dynamically modulating the external coupling rate $\kappa_\textrm{e}$, which results in the time-dependent variation of the total dissipation rate $\kappa$, as shown in Fig.\hyperref[fig1]{1(b)}. In general, for realistic pentacene maser platforms\cite{Oxborrow2012,Breeze2017,PhysRevApplied.14.064017}, the quantum battery is coupled with a high-$Q$ (low-$\kappa$) cavity continuously. The low dissipation of the cavity helps accumulate microwave photons and facilitates the stimulated emission of the quantum battery due to the presence of the spin-photon coupling. Thus, the work extraction occurs simultaneously with the charging process when the battery is not fully charged, leading to a relatively long-lasting and weak microwave emission. 

In contrast, by exploiting dissipation engineering, the work extraction can take place when the battery is fully charged, prior to which the cavity dissipation is set to a high value ($\kappa_\textrm{high}$) for a certain period $\tau_{1}$, so that the spin-photon coupling does not play important roles in the discharging of the battery. Moreover, the duration of work extraction can be manipulated by the period $\tau_{2}$ during which the cavity dissipation is switched to a low value ($\kappa_\textrm{low}$) that sustains for a period $\tau_\textrm{low}$ before recovery to $\kappa_\textrm{high}$. In principle, by compressing the interaction time between the fully charged quantum battery and cavity to a limited value (e.g., $\tau_2$), more battery energy can be employed for coherent microwave generation and the microwave pulse width can be substantially narrowed, leading to the enhanced output power. The complete cycle comprising $\tau_{1}$ and $\tau_{2}$ can be repetitive for generating microwave pulse trains. The above modulation approach is analogous to the so-called repetitive $Q$-switching technique\cite{481878} widely used for high-power pulsed lasers but with a distinct dissipation channel. Instead of introducing modulated \textit{internal} loss of the cavity, we focus on the control of \textit{external} coupling which directly connects to the microwave output characteristics.  

By taking practical feasibility into account, we investigate the effects of three different schemes of dissipation engineering in coherent microwave generation, which are called the \textit{instantaneous}, \textit{linear}, and \textit{sinusoidal} schemes. The difference arises from the manner of the switching between the $\kappa_\textrm{high}$ and $\kappa_\textrm{low}$ states, as shown in Fig.\hyperref[fig1]{1(b)}, since in reality, the switching normally takes a certain time that depends on the performance of the switching hardware components (e.g., the solid-state microwave switches) and the ringing behaviors of cavities. The instantaneous scheme demonstrates an ideal case that there are no transition phases between the $\kappa_\textrm{high}$ and $\kappa_\textrm{low}$ states, resulting in $\tau_2 = \tau_{\mathrm{low}}$. For the linear scheme, the switching is accomplished with the transition phases $\tau_{\mathrm{down}}$ and $\tau_{\mathrm{up}}$. During $\tau_{\mathrm{down}}$, the total dissipation reduces linearly according to the formula $\kappa=\kappa_{\mathrm{high}}- \left( \kappa_{\mathrm{high}} - \kappa_{\mathrm{low}}  \right) \times \Delta t / \tau_{\mathrm{down}}$, where $\Delta t$ is the time elapsed during the descent of $\kappa$ from its maximum value $\kappa_{\mathrm{high}}$ ($\Delta t \leq \tau_{\mathrm{down}}$). Subsequently, the dissipation increases during phase $\tau_{\mathrm{up}}$, following a trajectory that is a reverse of phase $\tau_{\mathrm{down}}$ ($\tau_{\mathrm{down}}=\tau_{\mathrm{up}}$). For the sinusoidal scheme, instead of implementing the linear transition phases, the dissipation varies sinusoidally, for instance, the dissipation in phase $\tau_{\mathrm{down}}$ varies according to the expression $\kappa=\kappa_{\mathrm{high}}- \left( \kappa_{\mathrm{high}} - \kappa_{\mathrm{low}}  \right) \times  \mathrm{sin}^2\left( \pi \Delta t / \left(2 \times \tau_{\textrm{down}} \right)  \right) $. Similarly, the variation in phase $\tau_{\mathrm{up}}$ shows an opposite trend to that in phase $\tau_{\mathrm{down}}$.

\section{MODEL}
We employ the reduced density operator $\hat{\rho}$ to represent the state of the metastable quantum battery, whose dynamical evolution is described by the following master equation\cite{Wu2024,Breeze2017}:
\begin{equation}\label{eq1}
\partial_\mathrm{t} \hat{\rho}=-\frac{i}{\hbar}\left[\hat{H}_{\mathrm{bat}}+\hat{H}_\mathrm{m}+\hat{H}_{\mathrm{m-bat}}, \hat{\rho}\right]+\mathcal{L}\left[ \hat\rho\right],
\end{equation}
where the Hamiltonian of the quantum battery $\hat{H}_{\mathrm{bat}}=\frac{1}{2}\hbar \omega_{\mathrm{35}} \sum_{k=1}^{N_{\mathrm{pen}}}\left(\hat{\sigma}_k^{\mathrm{55}}-\hat{\sigma}_k^{\mathrm{33}} \right) $. Here, $\hbar$, $k$, and $N_{\mathrm{pen}}$ are respectively the reduced Planck constant, the individual  pentacene molecule, and the total number of pentacene molecules. $\hat{\sigma}_k^{ij}=\left|i_k\right\rangle\left\langle j_k\right|$ is the spin transition operator. The Hamiltonian of the microwave cavity $\hat{H}_\mathrm{m}=\hbar \omega_\mathrm{m} \hat{a}^{\dagger} \hat{a}$, where $\hat{a}^{\dagger}$ $\left(\hat{a} \right)$ is the creation (annihilation) operator. The interaction between the quantum battery and the microwave cavity can be expressed by the Hamiltonian $\hat{H}_{\mathrm{m-bat}}=\hbar \sum_k g_{\mathrm{35}}\left(\hat{\sigma}_k^{\mathrm{53}} \hat{a}+\hat{a}^{\dagger} \hat{\sigma}_k^{\mathrm{35}}\right)$ with the coupling strength $g_{\mathrm{35}}$. $\mathcal{L}\left[ \hat\rho\right]$ is the Liouvillian, which accounts for the dissipative processes in the system. 

The full dynamics including all components of the Liouvillian can be found in \cor{Appendix A}, where the dissipation engineering is performed in the component describing the microwave photon loss in the cavity mode:
\begin{equation}\label{eq_dissipation}
\mathcal{L}_\textrm{m}\left[ \hat\rho\right] = \frac{\kappa}{2}\left[\left( n^{\mathrm{th}}_{\mathrm{m}} +1 \right)\mathcal{D}[\hat{a}] \hat{\rho}+n^{\mathrm{th}}_{\mathrm{m}} \mathcal{D}[\hat{a}^{\dagger}] \hat{\rho}\right],
\end{equation}
where the Lindblad superoperator $\mathcal{D}[\hat{\altmathcal{O}}]\hat{\rho}=2\hat{\altmathcal{O}}\hat{\rho}\hat{\altmathcal{O}}^\dagger-\hat{\altmathcal{O}}^\dagger\hat{\altmathcal{O}}\hat{\rho}-\hat{\rho}\hat{\altmathcal{O}}^\dagger\hat{\altmathcal{O}}$ and the thermal occupation number for temperature $T$ is defined by $n^{\mathrm{th}}_{\mathrm{m}}=[\textrm{exp}(\frac{\hbar\omega_\textrm{m}}{k_\textrm{B}T})-1]^{-1}$ with $k_\textrm{B}$ the Boltzmann constant, which is approximately 4000 at room temperature.

\begin{figure*}[htbp!]
	\includegraphics[width=0.98\textwidth]{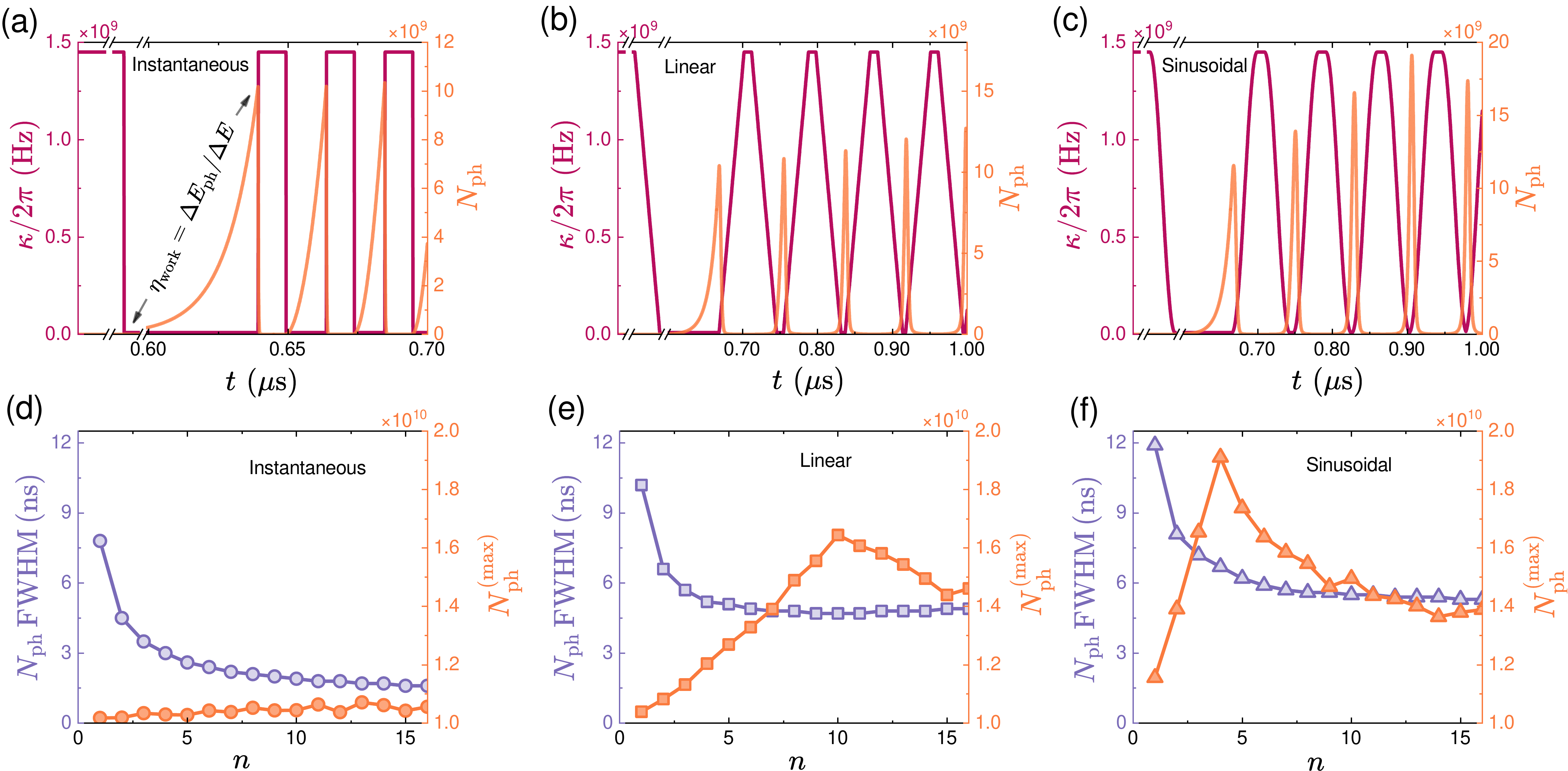}
	\caption{\label{fig2}  (a-c) Temporal evolution of the cavity total dissipation  rate $\kappa$ (left axis) and the microwave photon number $N_{\mathrm{ph}}$ (right axis) under three dissipation-engineering protocols.  The $\tau_\textrm{low}$ stage ends once the photon number reaches the threshold $N_{\mathrm{ph}} = 10^{10}$, after which the dissipation resets for the next cycle. The work extraction efficiency $\eta_{\mathrm{work}}$ is defined as the ratio between the photon energy change $\Delta E_{\mathrm{ph}}$ and the quantum battery energy change $\Delta E$ during each emission event, calculated from the start of the $\tau_2$ stage to the peak photon number. 
(d-f)  Evolution of the full width at half maximum (FWHM, left axis) and maximum photon number \cor{$N_{\mathrm{ph}}^{\mathrm{(max)}}$} (right axis) for successive (the $n$th) microwave pulses under the three modulation schemes. All simulations were performed with $\tau_1=10^{-8}$ s, $\kappa_{\mathrm{low}}/2\pi=9.55 \times 10^6$ Hz, $\tau_{\mathrm{down}}=2/ \kappa_{\mathrm{low}}$, and $\tau_{\mathrm{up}}=\tau_{\mathrm{down}}$. Additional system parameters are presented in Table $1$ of \cor{Appendix B}. Unless otherwise stated, all simulations adopt identical system parameters. 
}
\end{figure*}

By performing the mean-field calculations of the coupled differential equations derived from the master equation (\cor{see Appendix A}), the stored energy of the quantum battery can be expressed as $E(t)=N_{\mathrm{pen}}\hbar \omega_{\mathrm{35}} \langle \hat{\sigma}_\mathrm{1}^{\mathrm{55}}(t)\rangle$. Let $P_{\mathrm{ins}}(t)=\dot{E}(t)$ be the instantaneous power of the quantum battery. Meanwhile,
to quantify the process of the work extraction, the useful energy $E_\textrm{ph}(t)$ extracted from the quantum battery for coherent microwave generation can be expressed with the number of intra-cavity microwave photons $N_\textrm{ph}(t) = \langle \hat{a}^{\dagger}\hat{a}\rangle(t)$ as 
$\label{eq1_1}
E_\textrm{ph}(t)=N_{\mathrm{ph}}(t)\hbar\omega_{\mathrm{m}}.$ And the microwave output power is calculated using the formula
\begin{equation}\label{eq2}
P_{\mathrm{out}}(t)=N_{\mathrm{ph}}(t)\hbar\omega_{\mathrm{m}}\kappa k_\textrm{c}/(1+k_\textrm{c}),
\end{equation}
where $k_\textrm{c}$ is the coupling coefficient characterizing the output coupler of the cavity and defined as\cite{kajfez1998dielectric}
\begin{equation}\label{coupling}
    k_\textrm{c} = \frac{\kappa_\textrm{e}}{\kappa_{0}} = \frac{\kappa}{\kappa_{0}}-1.
\end{equation} 


\section{CONTROLLABLE MICROWAVE PULSE GENERATION}
By simulating the temporal evolution of the microwave photon number $N_{\mathrm{ph}}$ in the cavity under the three dissipation-engineering protocols, as shown in Fig.\hyperref[fig2]{2(a)-(c)}, we first verify the concept of dissipation engineering for controllable microwave pulse generation. As predicted, these schemes effectively modulate the coupling between the cavity field and the external environment, thereby shaping the dynamics of photon accumulation and release during each discharging cycle of the quantum battery.

In the instantaneous scheme, the abrupt switching of $\kappa$ simulates an ideal process, where energy stored in the quantum battery is suddenly extracted into the cavity mode and then immediately 'used' via the external coupling. This rapid energy extraction leads to sharply defined microwave pulses with well-controlled intensity and interval, which are indicated by the overlaps between the bursts of intra-cavity photon number ($N_\textrm{ph}$) and the modulation of dissipation in terms of the onset and termination positions. By contrast, in the linear and sinusoidal schemes, $\kappa$ changes continuously in time, producing a more gradual transition between the energy extraction and usage stages. As a result, the photon number continues to rise slightly even after reaching to the maximum obtained in the instantaneous scheme, reflecting a delayed energy extraction caused by the finite rate of dissipation engineering. This behavior mirrors the finite response of an oscillator under time-dependent dissipation engineering in realistic devices\cite{159520}.

Figs.\hyperref[fig2]{2(d)-(f)} summarize the evolution of the full width at half maximum (FWHM) and the maximum intra-cavity photon number [$N_{\mathrm{ph}}^{\mathrm{(max)}}$] of successive pulses. For all schemes, the FWHM narrows rapidly during the first few cycles and subsequently stabilizes. The reduction in pulse width arises from the growing initial photon number at the onset of each $\tau_2$ stage (see Fig.\hyperref[figS1]{S1} in \cor{Appendix B}). As the cavity stores emitted/residual photons between cycles, the field buildup becomes progressively faster in subsequent emissions, producing shorter and more intense pulses. Such effects have been proved by the seeding measurements in the room-temperature pulsed maser device\cite{Salvadori2017}. Eventually, as the growing self-seeded photons enable to saturate the maser transition, the FWHM tends to stabilize or even broaden, for instance, in the linear scheme.


Quantitatively, all three schemes compress the microwave pulses into the nanosecond regime, which is approximately three orders of magnitude narrower than the so far shortest maser pulse obtained at room temperature\cite{Salvadori2017}. Under the instantaneous modulation, the FWHMs of successive pulses decrease from $8$ ns to below $2$ ns, whereas the minimum FWHM around 5 ns is obtained in both linear and sinusoidal schemes. This difference can be attributed to the gradual modulations of $\kappa$ that allow partial energy leakage of the quantum battery in form of microwave emission during the phase $\tau_\textrm{down}$ and extend the emission tail during the phase $\tau_\textrm{up}$, thus limiting the achievable pulse compression.


The maximum intra-cavity photon number of each pulse exhibits fluctuations around the preset threshold $N_{\mathrm{ph}}=10^{10}$, which is imposed to prevent the FWHM broadening caused by excessive energy release per pulse. Since the threshold defines when the dissipation switching starts, the resulted $\tau_{\mathrm{low}}$ varies from cycle to cycle. For all schemes, $\tau_\textrm{down}$ of the first cycle typically lasts longer as shown in Fig.\hyperref[fig2]{2(a)-(c)} due to the initially less populated cavity, while later cycles proceed faster once residual photons act as a seed field. Among all schemes, the instantaneous scheme yields the most stable pulse trains, as its abrupt switching minimizes the influence of residual photons on the subsequent emission. In contrast, due to the prolonged spin-photon coupling during the smooth transition phases in the linear and sinusoidal modulations, the overall behavior of $N_{\mathrm{ph}}^{\mathrm{(max)}}$ shows a significant growth beyond the threshold with a non-monotonic trend. After reaching the maximums at respectively the 10th and 4th cycles of the linear and sinusoidal schemes, $N_{\mathrm{ph}}^{\mathrm{(max)}}$ of the subsequent pulses gradually drops to a quasi-stable value at $1.4\times10^{10}$. This occurs because, after the $10$th cycle, the linear scheme achieves the target photon number of $10^{10}$ already during phase $\tau_{\mathrm{down}}$, proceeds directly to phase $\tau_{\mathrm{up}}$, and does not undergo a complete phase $\tau_{\mathrm{up}}$, resulting in a decrease in photon number. A similar behavior is observed with the sinusoidal scheme, where the photon number reaches $10^{10}$ during phase $\tau_{\mathrm{down}}$ after the $4$th cycle. (see Fig.\hyperref[figS1]{S1(b)} in \cor{Appendix B})


\begin{figure*}
	\includegraphics[width=0.98\textwidth]{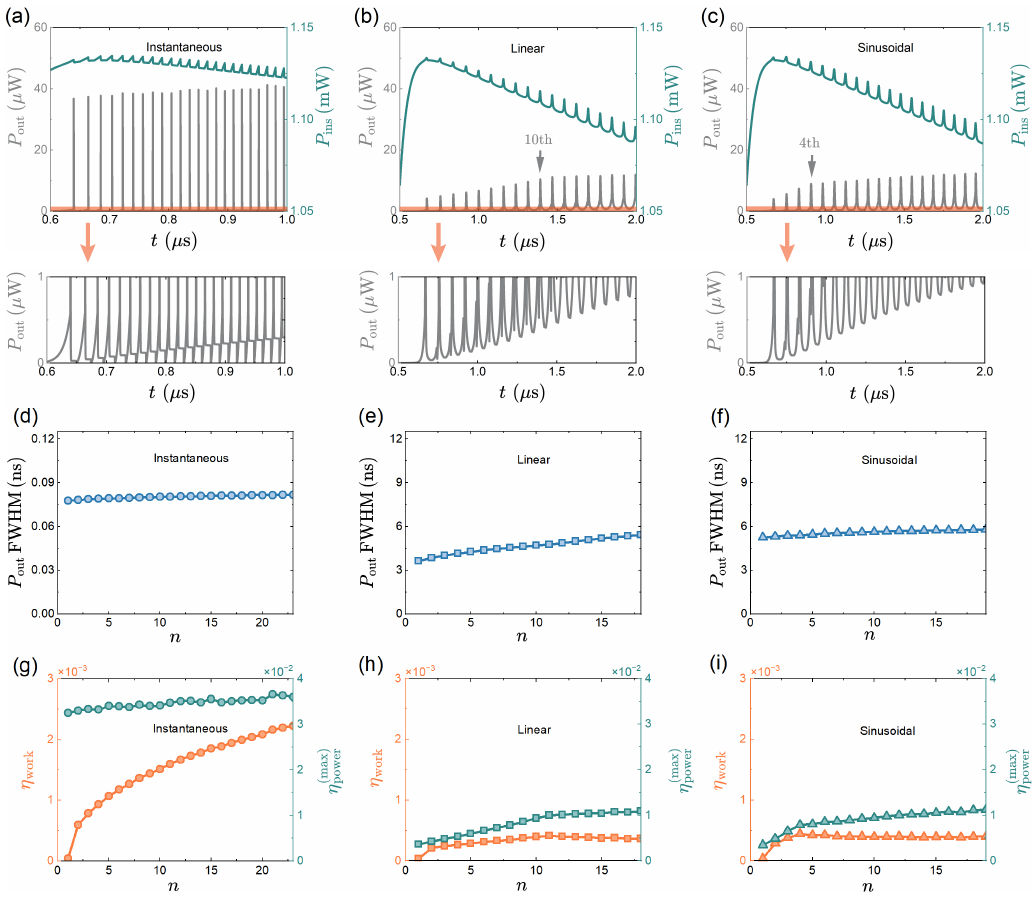}
	\caption{\label{fig3} (a-c) Temporal evolution of the output power $P_{\mathrm{out}}$ (left axis) and instantaneous power $P_{\mathrm{ins}}$ (right axis) under the three dissipation-engineering schemes. For clarity, the time scale in (a) is different from those in (b) and (c). The insets show magnified views of $P_{\mathrm{out}}$ within the organic shadow regimes. (d-f) Evolution of the FWHM of the output power $P_{\mathrm{out}}$ as well as (g-i) the work extraction efficiency $\eta_{\mathrm{work}}$ (left axis) and maximum power compression factor $\eta_{\mathrm{power}}^{(\mathrm{max})}$ (right axis, power compression factor at the maximum output power) for successive (the $n$th) microwave pulses under the three modulation schemes}
\end{figure*}
\section{MICROWAVE OUTPUT PERFORMANCE}

In addition to the analysis of the microwave generation inside the cavity, we also implement the simulation of the microwave output from the cavity based on Eqs.\hyperref[eq2]{(3)} and \hyperref[coupling]{(4)}. The temporal evolution of the output power $P_{\mathrm{out}}$ under the three dissipation modulation protocols is shown in Figs.\hyperref[fig3]{3(a)–(c)}. To help analyze the output performance, the instantaneous power $P_{\mathrm{ins}}$ indicating the time-dependence of the energy storage in the quantum battery is illustrated as well. The power stability of the output pulse trains obtained with the three schemes is similar to that of the intra-cavity photon number shown in Fig.\hyperref[fig2]{2}. The results demonstrate that the instantaneous scheme offers the possibility of achieving the most stable output pulses since the beginning of the dissipation modulation, whereas it takes several cycles for the other two schemes to stabilize the output. It is worth noting that the output pulses obtained with the sinusoidal modulation protocol start to become stable prior to those in the linear scheme, similar to the trend of reaching the peak of $N_{\mathrm{ph}}^{\mathrm{(max)}}$ in the linear and sinusoidal schemes, where the maximum photon numbers are established respectively in the $10$th and $4$th cycles as shown in Figs.\hyperref[fig2]{2(e)} and \hyperref[fig2]{2(f)}. Despite the decline of $N_{\mathrm{ph}}^{\mathrm{(max)}}$ after reaching the maximums in both schemes, the later-stage stability indicates the output power is more sensitive to the variation of the intra-cavity photon number when the cavity is less populated in the early stage. 

In terms of the magnitude of $P_{\mathrm{out}}$, with the non-optimal parameter setting, all schemes can lead to microwatt ($\mu$W)-level output power which is of the same order of magnitude as that of most experimentally demonstrated pentacene masers\cite{Salvadori2017,Oxborrow2012,Breeze2017,PhysRevApplied.14.064017,Wu:20,Breeze2015}. We note that the very recent experimental realizations show the possibility of reaching the milliwatt (mW) level with a much stronger optical pumping\cite{long2025feasibilityfreespacetransmissionusing,long2025lbandmilliwattroomtemperaturesolidstate}. Among them, the instantaneous scheme gives rise to the maximum output power, while its intra-cavity photon number is the lowest (just above the preset threshold) in Fig.\hyperref[fig2]{2}. The phenomenon can be explained by evaluating the instantaneous power $P_{\mathrm{ins}}$ shown in Figs.\hyperref[fig3]{3(a)–(c)}. Before the initialization of the modulation, the rise of $P_{\mathrm{ins}}$ under all schemes arises from the charging process. Note that, due to the denser pulses, the time scale used for plotting $P_{\mathrm{ins}}$ of the instantaneous scheme is different from those in the other two schemes so that the dramatic increase of $P_{\mathrm{ins}}$ is not shown in Fig.\hyperref[fig3]{3(a)}. Upon work extraction, it can be found that for the first cycle, all schemes demonstrate almost the same level of $P_{\mathrm{ins}}$, indicating the energy extracted from the battery for the microwave generation is similar. Because of the energy conservation, the more microwave photons coupled out of the cavity results in the fewer intra-cavity photons. Moreover, the instantaneous scheme shows the relatively stable $P_{\mathrm{ins}}$ over the pulse trains, where significant decrease is observed for the other two schemes in later cycles. According to the magnified views of Figs.\hyperref[fig3]{3(a)–(c)}, the power losses are attributed to the continuously increased photon leakage between cycles, where the transition regimes of the dissipation switching are present in both schemes.

In addition to the output power, the output pulse width is also crucial for practical applications. Figs.\hyperref[fig3]{3(d)–(f)} show the evolution of the FWHM of the output pulses for each scheme. Strikingly, by employing the ideal dissipation engineering, i.e., the instantaneous protocol, the output pulse widths can be drastically reduced to tens of picoseconds, which are two orders of magnitudes narrower compared to those obtained with the other protocols. Again, the broader output pulses at the level of nanoseconds result from the transition regimes in the linear and sinusoidal schemes. In contrast to the FWHM of $N_\textrm{ph}$ shown in  Fig.\hyperref[fig2]{2}, which presents a decrease trend in the initial pulses due to the buildup of cavity field, the FWHM of $P_\textrm{out}$ is relatively stable over the successive pulses and narrower. This may be attributed to dissipation engineering which largely isolates the output channels from the intra-cavity dynamics and thus the output pulse shape is governed by the modulation behaviors.


\begin{figure*}
	\includegraphics[width=0.98\textwidth]{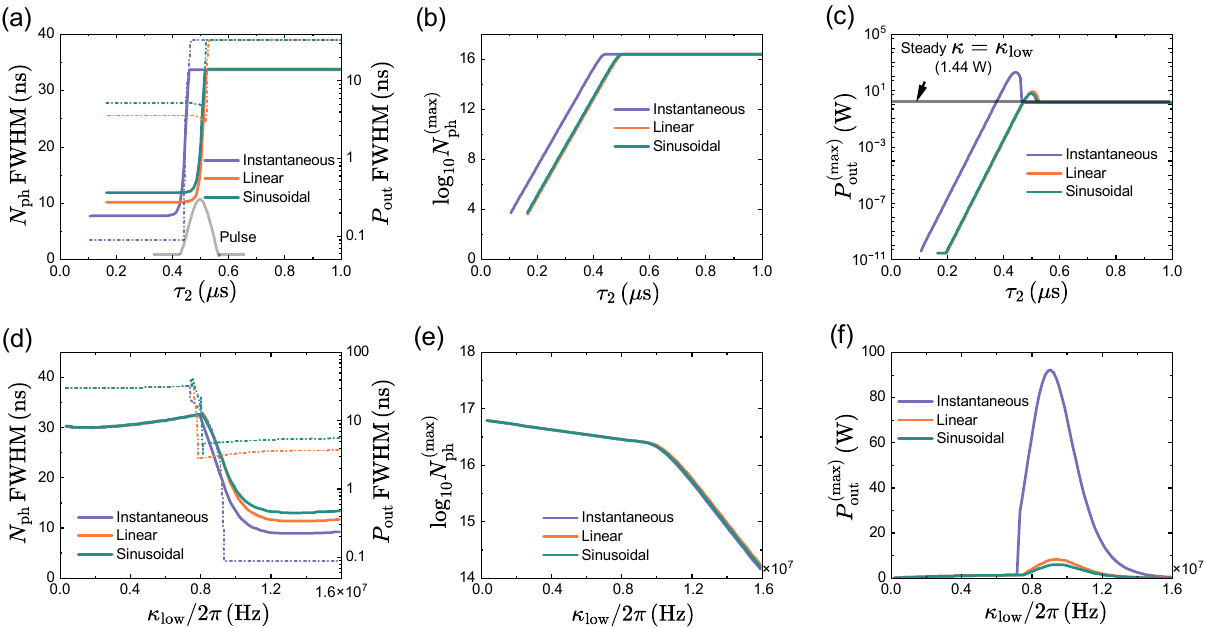}
	\caption{\label{fig4} Dependence of the FWHMs of the intra-cavity photon number $N_{\mathrm{ph}}$ (solid lines) and the microwave output $P_\textrm{ins}$ (dashed lines), the maximum intra-cavity photon number $N_{\mathrm{ph}}^{(\mathrm{max})}$, and the maximum output power $P_{\mathrm{out}}^{(\mathrm{max})}$ on (a-c) the dissipation modulation duration $\tau_2$ and 
    (d-f) the minimum dissipation rate $\kappa_{\mathrm{low}}$. In (a-c), the value of $\kappa_{\mathrm{low}}/2\pi$ is set to $9.55 \times 10^6$ Hz; In (d-f), the values of $\tau_2$ chosen for the instantaneous, linear and sinusoidal schemes are 440, 505 and 500 ns 
    respectively, which yield the maximum $P_{\mathrm{out}}^{(\mathrm{max})}$ in (c).}  
\end{figure*}

We further evaluate the efficiency of coherent microwave pulse generation with dissipation engineering. We define the work extraction efficiency $\eta_\textrm{work}=\Delta E_\textrm{ph}/\Delta E$ with the intra-cavity photon energy change $\Delta E_\textrm{ph}$ and the quantum battery energy change $\Delta E$ during each cycle. For consistency among the three schemes, the period chosen for the calculation starts from the onset of $\tau_2$ to the position where $N_{\mathrm{ph}}^{\mathrm{(max)}}$ is reached in each cycle, as illustrated in Fig.\hyperref[fig2]{2(a)}. As shown in Fig.\hyperref[fig3]{3(g)-(i)}, the instantaneous protocol achieves the highest $\eta_\textrm{work}$, benefiting from the abrupt release of stored energy that against the non-radiative losses\cite{SIXL197021} of the pentacene triplet spins. In contrast, the linear and sinusoidal modulations yield similar, but lower efficiencies because of the partial loss of the battery energy during the gradual modulation of $\kappa$. In the first few cycles, $\eta_\textrm{work}$ remains small in all schemes due to the initial low photon number at the start of the $\tau_2$ stage, where the ohmic and dielectric dissipation\cite{kajfez1998dielectric} of the cavity, constituting the internal dissipation ($\kappa_0$), dominates over the stimulated emission. As the cavity field builds up after several cycles, $\eta_\textrm{work}$ increases and eventually saturates following the similar trend of $N_{\mathrm{ph}}^{\mathrm{(max)}}$. We note that with the current parameter setting, the overall $\eta_\textrm{work}$ of the order of $10^{-3}$ is rather low, of which the optimization will be presented in the next section.

Moreover, we investigate the power compression factor $\eta_\textrm{power}$ of different modulation protocols, which is a quantity employed in the field of (sub)nanosecond microwave pulse generation\cite{10857811,PhysRevLett.92.118301}, to characterize how effectively a long-duration, low-power input pulse is converted into a short-duration, high-power output pulse. In our case, there are no microwave pulses injected in the cavity, instead, the pulsed discharging from the quantum battery can be considered as the 'input' source, linking the microwave output power $P_\textrm{out}$ to the instantaneous power $P_\textrm{ins}$ of the quantum battery. Figs.\hyperref[fig3]{3(g)-(i)} illustrate the maximum power compression factors $\eta_{\mathrm{power}}^{(\mathrm{max})}$ in the three schemes that is calculated by the ratio of $P_{\mathrm{out}}$ to $P_{\mathrm{ins}}$ at the maximum output power in each cycle shown in Figs.\hyperref[fig3]{3(a)-(c)}.  Similar to $\eta_\textrm{work}$, the instantaneous scheme exhibit the highest power compression factor, but the magnitude is still far below unity ($\sim10^{-2}$), indicating that although the power is temporally compressed, the efficiency of converting the battery' energy into high-power output remains modest.

\section{OPTIMIZATION OF DISSIPATION ENGINEERING}

Finally, we analyze the optimization of dissipation engineering by tuning the modulation duration $\tau_2$ and the minimum dissipation rate $\kappa_{\mathrm{low}}$. Fig.\hyperref[fig4]{4} shows the impact of these two parameters on the microwave photon generation and output characteristics of the first pulse produced under the three schemes. As shown in Fig.\hyperref[fig4]{4(a)}, both FWHMs of the temporal evolutions of $N_\textrm{ph}$ and $P_\textrm{out}$ remain nearly constant at short $\tau_2$, then broaden sharply before stabilizing near their maximums. We find the abrupt broadening starts when the intra-cavity photon is accumulated close to the maximum. Therefore, the values of $\tau_2$ (approximately 450 ns 
in the instantaneous schemes and 500 ns
in the other two schemes) leading to the turning points of the FWHMs match those observed in Fig.\hyperref[fig4]{4(b)}, where the saturation of the maximum intra-cavity photon numbers $N_{\mathrm{ph}}^{(\mathrm{max})}$ $\sim10^{16}$ start to occur after a linear growth. Despite the similar trend of the FWHM between $N_\textrm{ph}$ and $P_\textrm{out}$, the variation of the magnitude is distinct. For the intra-cavity photons, the pulse widths obtained under all schemes increase from near 10 ns 
to about 35 ns. 
In contrast, for the output, the pulse widths can change substantially by three orders of magnitude (tens of ps 
to more than 10 ns 
) in the instantaneous scheme and maximally one order of magnitude 
in the other two schemes. \cor {Note that factors such as the detuning between the generated microwave photons and cavity mode are not considered here. In practice, an nanosecond microwave pulse in the frequency band of GHz is feasible, however, the simulated ps-scale FWHM including both the results obtained before and after the optimization, shown in Fig.\hyperref[fig3]{3} and  Fig.\hyperref[fig4]{4}, implying super-broadband distributions of the photon frequencies, leads to a vast off-resonance condition. Thus, the ps-level linewidths are only of theoretical significance, while the experimental demonstration would be largely restricted by the cavity mode linewidth.}

The maximum output power $P_{\mathrm{out}}^{(\mathrm{max})}$ can also be dramatically affected by $\tau_2$ as shown in Fig.\hyperref[fig4]{4(c)}. The highest $P_{\mathrm{out}}^{(\mathrm{max})}$ occur with the values of $\tau_2$ chosen to be close to the ones giving rise to the turning points obtained in Figs.\hyperref[fig4]{4(a)} and \hyperref[fig4]{(b)}, which are approaching 100 W using the instantaneous protocol, and 10 W for the other protocols. It is worth noting that, due to the sharp transitions of the FWHM of $P_\textrm{out}$ in all schemes, $P_{\mathrm{out}}^{(\mathrm{max})}$ is so sensitive to $\tau_2$ that a slight extension of $\tau_2$ beyond the optimum can significantly degrade the output power to 1.44 W, which is the same as that obtained without the dissipation modulation. We also find once $\tau_2$ is prolonged to specific values, the further increase of $\tau_2$ can no longer alter the intra-cavity photon characteristics and microwave output performance, and all three schemes converge to identical results.


Figs.\hyperref[fig4]{4(d-f)} further show the optimization of the minimum dissipation rate $\kappa_{\mathrm{low}}$ based on the optimal $\tau_2$ obtained in Fig.\hyperref[fig4]{4(c)} which leads to the highest microwave output power. Therefore, the values of $\tau_2$ are chosen to be 440, 505 and 500 ns 
for the instantaneous, linear and sinusoidal schemes respectively.   
As $\kappa_{\mathrm{low}}/2\pi > 9.55 \times 10^6 $ Hz, the system gradually fails to generate a complete pulse (see Fig.\hyperref[figS2]{S2} and Fig.\hyperref[figS3]{S3}   in Appendix B), leading to a rapid decline in the maximum intra-cavity photon number shown in Fig.\hyperref[fig4]{4(e)}. Such a trend is identical for all three schemes. Meanwhile, the FWHM of $N_{\mathrm{ph}}$ decreases by approximately one-third of its initial value, 30 ns 
to a stable value around 10 ns 
when $\kappa_\textrm{low}/2\pi$ is increased above $7.96\times10^6$ Hz (see Fig.\hyperref[fig4]{4(d)}). With the increase of $\kappa_\textrm{low}$, the FWHM of the output, initially at around 30 ns, 
decreases to 4$\sim$6 ns 
for the linear and sinusoidal schemes, and further down to below 0.1 ns 
for the instantaneous scheme. This agrees with the dependence of the FWHM on $\tau_2$, consolidating the fact that the instantaneous scheme can realize much stronger temporal confinement of the output pulse. 
\begin{figure*}[htb!]
	\includegraphics[width=0.75\textwidth]{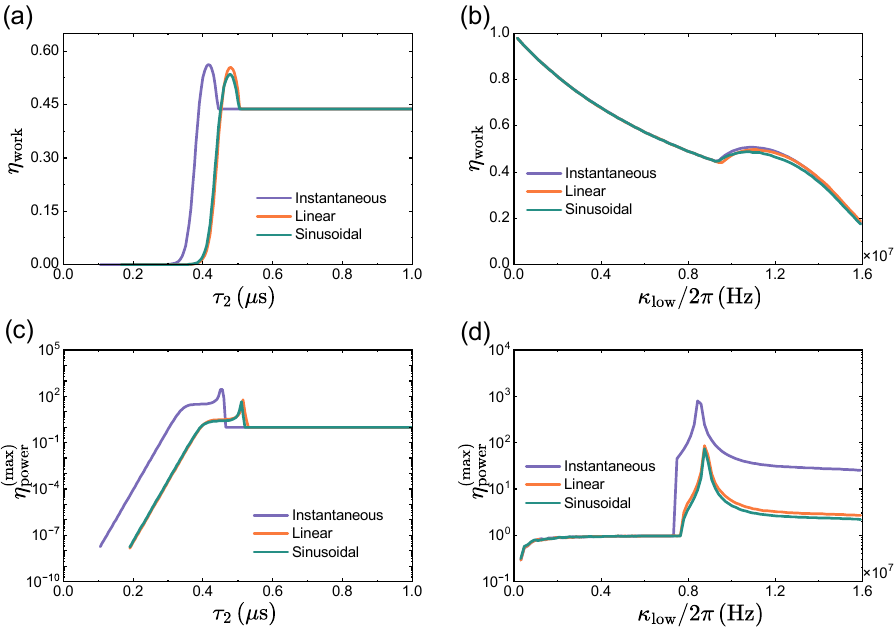}
	\caption{\label{fig5}  (a,b) The work extraction efficiency $\eta_{\mathrm{work}}$ and (c,d) maximum power compression factor $\eta_{\mathrm{power}}^{(\mathrm{max})}$ depending on $\tau_2$ and $\kappa_{\mathrm{low}}$. In (b) and (d), $\tau_2$ of the instantaneous, linear, and sinusoidal schemes are 440, 505 and 500 ns 
    respectively.}
\end{figure*}
As shown in Fig.\hyperref[fig4]{4(f)}, the output peak powers obtained with the three schemes are the same as those shown in Fig.\hyperref[fig4]{4(c)}. The deviations of $\kappa_\textrm{low}$ from its optimal value ($\kappa_{\mathrm{low}}/2\pi=9.55 \times 10^6$ Hz) lead to a reduction of the peak power in all schemes. Since the internal dissipation is fixed in this study, the smaller $\kappa_\textrm{low}$ indicates the better energy accumulation inside the cavity while a large $\kappa_\textrm{low}$ facilitates the energy extraction to external environment. The existence of the optimal $\kappa_\textrm{low}$ implies the critical balance between the energy accumulation and external coupling required to optimize the output power of the coherent microwave generation in the quantum battery.

\begin{figure}[tb!]
	\includegraphics[width=0.45\textwidth]{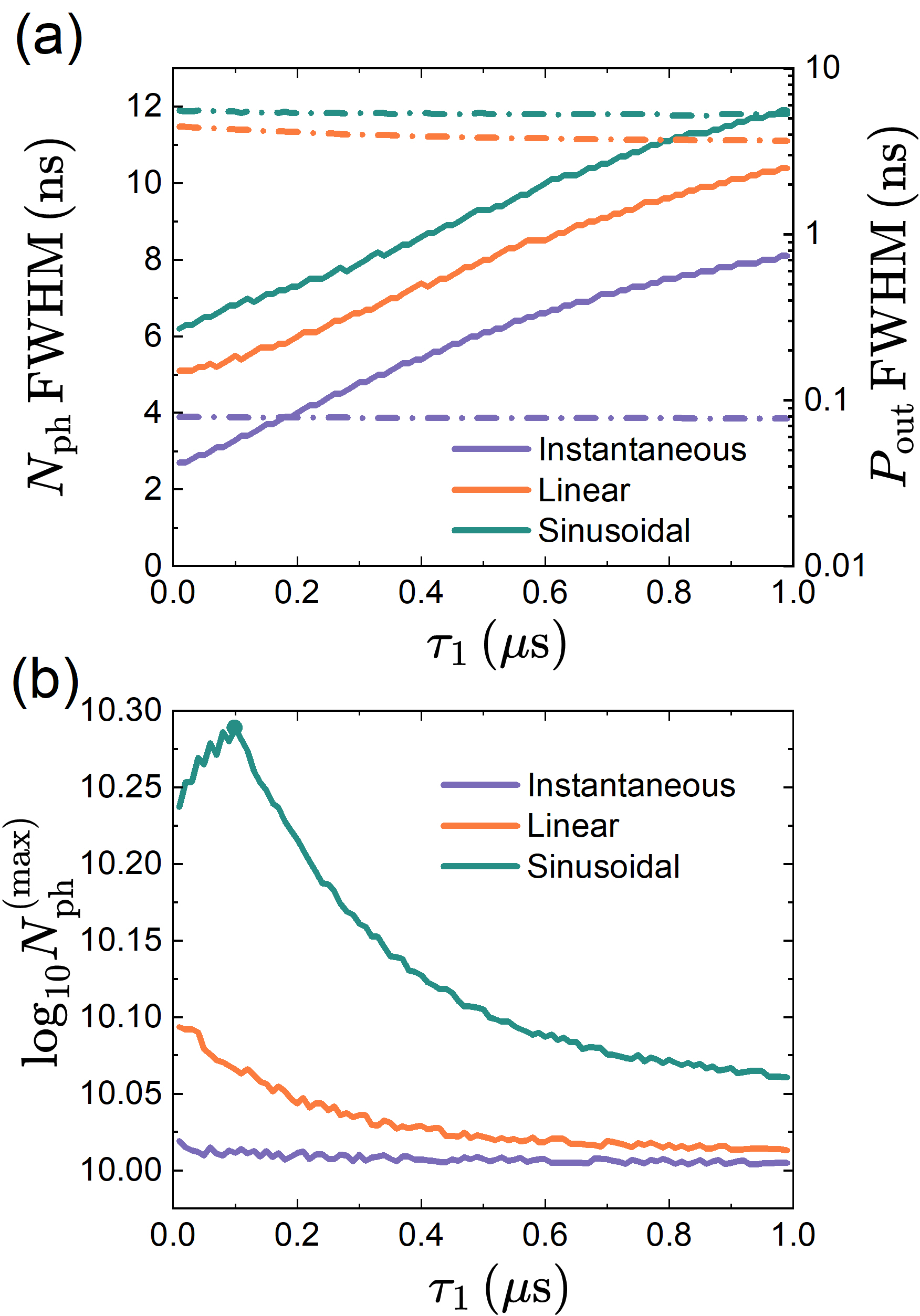}
	\caption{\label{fig6} (a) Variations of the FWHMs of $N_{\mathrm{ph}}$ (solid lines) and$P_{\mathrm{out}}^{(\mathrm{max})}$ (dashed lines) of the fifth output pulse with the delay time $\tau_1$. (b) The influence of $\tau_1$ on the maximum photon number $N_{\mathrm{ph}}^{(\mathrm{max})}$ of the fifth pulse.}
\end{figure}

With the above optimization, we can evaluate the work extraction efficiency $\eta_{\mathrm{work}}$ and maximum power compression factor $\eta_{\mathrm{power}}^{(\mathrm{max})}$ in the optimal conditions. 
Fig.\hyperref[fig5]{5(a)} reveals that for all three schemes, beyond specific thresholds of $\tau_2$, $\eta_{\mathrm{work}}$ increases rapidly up to about 0.6, which is two orders of magnitude improvement compared to the non-optimized results shown in Fig.\hyperref[fig2]{2}, and then stabilizes at 0.45. The highest $\eta_{\mathrm{work}}$ achievable for all schemes are similar and the corresponding $\tau_2$ values agree with the ones leading to the maximum intra-cavity photon numbers as discussed in Fig.\hyperref[fig4]{4}. It can be understood according to the definition of $\eta_\textrm{work}$, of which the denominator $\Delta E_\textrm{ph}$ associated with the intra-cavity photon energy is maximized. On the other hand, reducing $\kappa_{\mathrm{low}}$ can improve $\eta_\textrm{work}$ as shown in Fig.\hyperref[fig5]{5(b)}. It can be explained that as $\kappa_{\mathrm{low}}$ decreases, the output channel is gradually closed, so that most photon energy extracted from the quantum battery can be well accumulated in the cavity, and once the stimulated emission dominates over the internal dissipation of the cavity, the highest $\eta_{\mathrm{work}}$ can approach the unity. Nevertheless, for applications, it is undesired to always trap the photons inside the cavity, thus a certain level of dissipation to external environment is required. Despite the reduction in $\eta_\textrm{work}$, the local minium of $\eta_\textrm{work}$ appears at the optimal $\kappa_\textrm{low}/2\pi$ around $9.55 \times 10^6 $ Hz which leads to the highest output power [see Fig.\hyperref[fig4]{4(f)}]. 

The most striking improvements due to the optimization are found in the power compression factor. Compared to the ones obtained in  Fig.\hyperref[fig3]{3}, the power compression factors of all schemes can be enhanced by four to five orders of magnitude, as shown in Fig.\hyperref[fig5]{5(c)} and \hyperref[fig5]{5(d)}. The maximum power compression factors realized by the optimization are $10^{3}$ in the instantaneous scheme, and $10^{2}$ in the other schemes. So far, most high-power microwave pulse compressors possess the power compression factors below $10^{2}$\cite{10857811,PhysRevLett.92.118301,7027876,10530991} except a time-reversal microwave pulse-compression cavity that can achieve a power compression factor of about 133\cite{9130060}. Note that the values of $\tau_2$ and $\kappa_\textrm{low}$ giving rise to optimal $\eta_{\mathrm{power}}^{(\mathrm{max})}$ slightly differs from the ones leading to the highest output power. And significant fluctuations of $\eta_{\mathrm{power}}^{(\mathrm{max})}$ can be caused by the deviations of the optimal $\tau_2$ and $\kappa_{\mathrm{low}}$. The above observations can be attributed to the abnormal dependence of the instantaneous power $P_{\mathrm{ins}}$ of the quantum battery on $\tau_2$ and $\kappa_{\mathrm{low}}$, where substantial reduction of $P_{\mathrm{ins}}$ is observed near the the optimal $\tau_2$ and $\kappa_{\mathrm{low}}$ (see Fig.\hyperref[figS4]{S4} in \cor{Appendix B}).

Furthermore, we investigate the impact of the delay time $\tau_1$ on the coherent microwave generation as shown in Fig.\ref{fig6}. Given that the variations in $\tau_1$ across the three schemes do not have a significant effect on the first pulse, we select the fifth pulse for comparison. Evidently, a shorter $\tau_1$ corresponds to a higher initial photon number at the onset of modulation time $\tau_2$. This is because during phase $\tau_1$, where the dissipation rate $\kappa=\kappa_{\mathrm{high}}$, a longer duration results in a greater reduction in the photon number. As illustrated in \hyperref[fig6]{6(a)}, the shorter delays ($\tau_1$) tend to reduce the FWHM of the intra-cavity photon burst, while expanding $\tau_1$ will slightly shorten the FWHM of the output pulse. In Fig.\hyperref[fig6]{6(b)}, the reduction of $\tau_1$ does not significantly affect the maximum photon number under the instantaneous scheme, however, it results in a gradual increase in the maximum intra-cavity photon number under the linear scheme. In the sinusoidal scheme, as the parameter $\tau_1$ decreases, the maximum intra-cavity photon number initially increases until the value of $\tau_1$ falls below 100 ns, after which it begins to decrease. This behavior can be attributed to the fact that when $\tau_1$ becomes excessively small, the dissipation rate fails to fully undergo the complete sequence of processes $\tau_{\mathrm{down}}$ and $\tau_{\mathrm{up}}$, thereby missing a portion of the mechanism responsible for photon generation. Except the FWHMs and the maximum intra-cavity microwave photon number, we find the impact of $\tau_1$ on other parameters, e.g., the work extraction efficiency and power compression factor is negligible and is thus not presented here.



\section{DISCUSSION AND CONCLUSION}
In summary, we introduce the concept of dissipation engineering as a means to dynamically control the pathways by which stored quantum battery energy is converted into propagating microwave photons. Analogous to \textit{Q}-switching\cite{McClung:62} in laser physics and to engineered reservoirs in CQED systems, dissipation engineering separates the charging and discharging stages: during charging, the system suppresses emission to accumulate energy in long-lived metastable states; during discharging, the coupling to the output channel is rapidly enhanced, enabling a burst of coherent, Purcell-enhanced microwave emission into the load. With optimal dissipation engineering, we successfully enhance the work extraction efficiency of the quantum battery by more than two orders of magnitude, and achieve watt-level maser output pulses with adjustable power and width. The  FWHMs of the intra-cavity microwave photon burst and the output pulse can both reach the nanosecond level. In extreme cases, the FWHM of the output power can reach the order of sub-nanoseconds when $\kappa$ undergoes an instantaneous change, while the feasibility of achieving such a short pulse experimentally is to be further investigated due to the potential mismatch between the broadband photon frequency distribution and the fixed cavity mode. Dissipation engineering near the pulse peak can significantly alter the FWHM and output power. These results demonstrate that properly timed dissipation modulation near the energy-release peak enables simultaneous control of pulse duration and intensity, thereby realizing highly efficient and temporally compressed coherent microwave emission. Especially, the power compression factors of this quantum battery is expected to surpass the state-of-the-art microwave pulse compression technique\cite{9130060}. Among the three dissipation schemes compared, instantaneous scheme can achieve the smallest FWHM, the highest output power, and the highest work extraction efficiency.  

From the standpoint of quantum thermodynamics, the presented strategy reflects a controlled modulation of entropy flow—temporarily isolating the battery from its dissipative environment, then selectively opening a channel that maximizes useful work extraction. By embedding this principle into the architecture of solid-state quantum batteries, our work provides a pathway to overcome the intrinsic power limitations of maser systems and to establish a room-temperature platform for high-power coherent microwave generation. This platform may also be applicable for the microwave analogy of supercontinuum generation\cite{BrèsDellaTorreGrassaniBraschGrilletMonat+2023+1199+1244} considering the broadband output of the ultra-short pulse simulated in the instantaneous scheme. Rooted on the modulated spin-photon coupling revealed in this study, we can envision another route of dissipation engineering based on the detuning strategy, by which the spin-coupling can also be tuned with the Zeeman/Stark shift of the molecular spin states or the dielectric constant of the cavity\cite{Shlafman:10} for high-power microwave output\cite{Ziemkiewicz:19}.

For future experimental validation, dissipation engineering can be implemented by connecting the microwave cavity to an actively controlled load for varying the external energy dissipation channel. 
The dissipation rate is then adjusted by controlling the coupling between the cavity and the load via a fast switch as employed in the previous cavity cooling experiments\cite{chen2024overcomingthermalnoiselimitmicrowave}. By monitoring the generated microwave photons through a detector type log amplifiers and a comparator\cite{10205706}, the coupling strength between the microwave cavity and the load is regulated in real time to obtain microwave pulses with a given threshold.

\textit{Acknowledgements.---}The work is supported by the National Natural Science Foundation of China (Grant No. 12204040 and 12574382), and the Beijing Institute of Technology Research Fund Program for Young Scholars (Grant No. XSQD-6120230016).


\appendix
\section{MASTER EQUATION}

\setcounter{figure}{0}
\setcounter{equation}{0}
\renewcommand{\thefigure}{S\arabic{figure}}
\label{A}
The complete dynamics of the quantum battery are described by\cite{Wu2024,Breeze2017}:
\begin{equation*}\label{eq:eq1}
\begin{aligned}
 \partial_\mathrm{t} \hat{\rho}=&-\frac{i}{\hbar}\left[\hat{H}_{\mathrm{bat}}+\hat{H}_\mathrm{m}+\hat{H}_{\mathrm{m-bat}}, \hat{\rho}\right] \\
& +\frac{1}{2}\xi \sum_\mathrm{k} \mathcal{D}\left[\hat{\sigma}_\mathrm{k}^{\mathrm{21}}\right] \hat{\rho}+\frac{1}{2}\left(\xi+k_{\mathrm{sp}}\right) \sum_\mathrm{k} \mathcal{D}\left[\hat{\sigma}_\mathrm{k}^{\mathrm{12}}\right] \hat{\rho} \\
& +\sum_\mathrm{k}\left(\sum_{\mathrm{i=3,4,5}}\frac{1}{2} k_{\mathrm{2i}} \mathcal{D}\left[\hat{\sigma}_\mathrm{k}^{\mathrm{i2}}\right] \hat{\rho}+\sum_{\mathrm{i=3,4,5}}\frac{1}{2} k_{\mathrm{i1}} \mathcal{D}\left[\hat{\sigma}_\mathrm{k}^{\mathrm{1i}}\right] \hat{\rho}\right) \\
& +\sum_\mathrm{k} \sum_{\mathrm{i, j=3,4,5 ; i \neq j}}\left(\frac{1}{2}k_{\mathrm{ij}} \mathcal{D}\left[\hat{\sigma}_\mathrm{k}^{\mathrm{ji}}\right] \hat{\rho}\right) \\
& +\sum_\mathrm{k} \sum_{\mathrm{i, j=3,4,5 ; i < j}}\left(\frac{1}{4} \chi_{\mathrm{ij}} \mathcal{D}\left[\hat{\sigma}_\mathrm{k}^{\mathrm{jj}}-\hat{\sigma}_\mathrm{k}^{\mathrm{ii}}\right] \hat{\rho}\right) \\
& +\frac{\kappa}{2}\left[\left( n^{\mathrm{th}}_{\mathrm{m}} +1 \right)\mathcal{D}[\hat{a}] \hat{\rho}+n^{\mathrm{th}}_{\mathrm{m}} \mathcal{D}[\hat{a}^{\dagger}] \hat{\rho}\right],
\end{aligned}
\end{equation*}
with the Lindblad superoperator $\mathcal{D}[\hat{\altmathcal{O}}]\hat{\rho}=2\hat{\altmathcal{O}}\hat{\rho}\hat{\altmathcal{O}}^\dagger-\hat{\altmathcal{O}}^\dagger\hat{\altmathcal{O}}\hat{\rho}-\hat{\rho}\hat{\altmathcal{O}}^\dagger\hat{\altmathcal{O}}$. The second through sixth lines sequentially present all components of the Lindblad superoperator $\mathcal{L}\left[ \rho\right]$ introduced in the main text, namely the optical pumping and the spontaneous emission, the inter-system crossing,  the spin-lattice relaxation, the dephasing process, and the thermal emission or excitation of the microwave cavity. The optical pumping employed for the quantum battery is a $2\times 10^4$ W laser. (See Table.\ref{table1} for parameter definitions.) The derived system of equations necessary for analyzing the discharge process is characterized by:
\begin{equation*}
\begin{aligned}
\frac{d}{d t} \langle  \hat{\sigma}_\mathrm{1}^{\mathrm{11}}  \rangle= & -\xi \langle  \hat{\sigma}_\mathrm{1}^{\mathrm{11}}  \rangle+\left(\xi+k_{\mathrm{sp}}\right) \langle  \hat{\sigma}_\mathrm{1}^{\mathrm{22}}  \rangle  +k_{\mathrm{31}} \langle  \hat{\sigma}_\mathrm{1}^{\mathrm{33}}  \rangle \\
&+k_{\mathrm{41}} \langle  \hat{\sigma}_\mathrm{1}^{\mathrm{44}}  \rangle+k_{\mathrm{51}} \langle  \hat{\sigma}_\mathrm{1}^{\mathrm{55}}  \rangle \\
\frac{d}{d t} \langle  \hat{\sigma}_\mathrm{1}^{\mathrm{22}}  \rangle= & \xi \langle  \hat{\sigma}_\mathrm{1}^{\mathrm{11}}  \rangle-\left(\xi+k_{\mathrm{sp}}\right) \langle  \hat{\sigma}_\mathrm{1}^{\mathrm{22}}  \rangle -k_{\mathrm{23}} \langle  \hat{\sigma}_\mathrm{1}^{\mathrm{22}}  \rangle-k_{\mathrm{24}} \langle  \hat{\sigma}_\mathrm{1}^{\mathrm{22}}  \rangle \\
&-k_{\mathrm{25}} \langle  \hat{\sigma}_\mathrm{1}^{\mathrm{22}}  \rangle \\
\frac{d}{d t} \langle  \hat{\sigma}_\mathrm{1}^{\mathrm{33}}  \rangle=& i g_{\mathrm{35}}\left(\langle  \hat{\sigma}_\mathrm{1}^{\mathrm{53}} \hat{a} \rangle - \langle \hat{a}^{\dagger} \hat{\sigma}_\mathrm{1}^{\mathrm{35}}  \rangle\right)+k_{\mathrm{23}} \langle  \hat{\sigma}_\mathrm{1}^{\mathrm{22}}  \rangle 
-k_{\mathrm{31}} \langle  \hat{\sigma}_\mathrm{1}^{\mathrm{33}}  \rangle  \\
&-k_{\mathrm{34}} \langle  \hat{\sigma}_\mathrm{1}^{\mathrm{33}}  \rangle +k_{\mathrm{43}} \langle  \hat{\sigma}_\mathrm{1}^{\mathrm{44}}  \rangle-k_{\mathrm{35}} \langle  \hat{\sigma}_\mathrm{1}^{\mathrm{33}}  \rangle +k_{\mathrm{53}} \langle  \hat{\sigma}_\mathrm{1}^{\mathrm{55}}  \rangle \\
\frac{d}{d t} \langle  \hat{\sigma}_\mathrm{1}^{\mathrm{44}}  \rangle= & k_{\mathrm{24}} \langle  \hat{\sigma}_\mathrm{1}^{\mathrm{22}}  \rangle-k_{\mathrm{41}} \langle  \hat{\sigma}_\mathrm{1}^{\mathrm{44}}  \rangle +k_{\mathrm{34}} \langle  \hat{\sigma}_\mathrm{1}^{\mathrm{33}}  \rangle  -k_{\mathrm{43}} \langle  \hat{\sigma}_\mathrm{1}^{\mathrm{44}}  \rangle \\
&-k_{\mathrm{45}} \langle  \hat{\sigma}_\mathrm{1}^{\mathrm{44}}  \rangle+k_{\mathrm{54}} \langle  \hat{\sigma}_\mathrm{1}^{\mathrm{55}}  \rangle \\
\frac{d}{d t} \langle  \hat{\sigma}_\mathrm{1}^{\mathrm{55}}  \rangle=&i g_{\mathrm{35}}\left( \langle \hat{a}^{\dagger} \hat{\sigma}_\mathrm{1}^{\mathrm{35}}  \rangle-\langle  \hat{\sigma}_\mathrm{1}^{\mathrm{53}}\hat{a}   \rangle \right)+k_{\mathrm{25}} \langle  \hat{\sigma}_\mathrm{1}^{\mathrm{22}}  \rangle 
-k_{\mathrm{51}} \langle  \hat{\sigma}_\mathrm{1}^{\mathrm{55}}  \rangle  \\
& +k_{\mathrm{35}} \langle  \hat{\sigma}_\mathrm{1}^{\mathrm{33}}  \rangle -k_{\mathrm{53}} \langle  \hat{\sigma}_\mathrm{1}^{\mathrm{55}}  \rangle+k_{\mathrm{45}} \langle  \hat{\sigma}_\mathrm{1}^{\mathrm{44}}  \rangle -k_{\mathrm{54}} \langle  \hat{\sigma}_\mathrm{1}^{\mathrm{55}}  \rangle \\
\frac{d}{d t} \langle \hat{a}^{\dagger} \hat{a}  \rangle =&N i g_{\mathrm{35}}\left(\langle  \hat{\sigma}_\mathrm{1}^{\mathrm{53}} \hat{a} \rangle - \langle \hat{a}^{\dagger}  \hat{\sigma}_\mathrm{1}^{\mathrm{35}}  \rangle\right)-\kappa \langle \hat{a}^{\dagger} \hat{a}  \rangle 
\end{aligned}
\end{equation*}
\begin{equation*}
\begin{aligned}
\frac{d}{d t} \langle \hat{a}^{\dagger}  \hat{\sigma}_\mathrm{1}^{\mathrm{35}}  \rangle = & i \omega_\mathrm{m}  \langle \hat{a}^{\dagger}  \hat{\sigma}_\mathrm{1}^{\mathrm{35}}  \rangle-i \omega_{\mathrm{35}}  \langle \hat{a}^{\dagger}  \hat{\sigma}_\mathrm{1}^{\mathrm{35}}  \rangle + i g_{\mathrm{35}}(N-1)  \langle \hat{\sigma}_\mathrm{1}^{\mathrm{53}}  \hat{\sigma}_\mathrm{2}^{\mathrm{35}}  \rangle \\
&+ i g_{\mathrm{35}}\left[\left(1+\langle \hat{a}^{\dagger} \hat{a}  \rangle\right) \langle  \hat{\sigma}_\mathrm{1}^{\mathrm{55}}  \rangle-\langle \hat{a}^{\dagger} \hat{a}  \rangle \langle  \hat{\sigma}_\mathrm{1}^{\mathrm{33}}  \rangle \right]\\
& -\frac{1}{2} k_{\mathrm{31}}  \langle \hat{a}^{\dagger}  \hat{\sigma}_\mathrm{1}^{\mathrm{35}}  \rangle -\frac{1}{2} k_{\mathrm{51}}  \langle \hat{a}^{\dagger}  \hat{\sigma}_\mathrm{1}^{\mathrm{35}}  \rangle-\frac{1}{2} k_{\mathrm{34}}  \langle \hat{a}^{\dagger}  \hat{\sigma}_\mathrm{1}^{\mathrm{35}}  \rangle \\
&-\frac{1}{2} k_{\mathrm{35}}  \langle \hat{a}^{\dagger}  \hat{\sigma}_\mathrm{1}^{\mathrm{35}}  \rangle-\frac{1}{2} k_{\mathrm{53}}  \langle \hat{a}^{\dagger}  \hat{\sigma}_\mathrm{1}^{\mathrm{35}}  \rangle -\frac{1}{2} k_{\mathrm{54}}  \langle \hat{a}^{\dagger}  \hat{\sigma}_\mathrm{1}^{\mathrm{35}}  \rangle\\
&  -\frac{1}{4} \chi_{\mathrm{34}}  \langle \hat{a}^{\dagger}  \hat{\sigma}_\mathrm{1}^{\mathrm{35}}  \rangle-\chi_{\mathrm{35}}  \langle \hat{a}^{\dagger}  \hat{\sigma}_\mathrm{1}^{\mathrm{35}}  \rangle -\frac{1}{4} \chi_{\mathrm{45}}  \langle \hat{a}^{\dagger}  \hat{\sigma}_\mathrm{1}^{\mathrm{35}}  \rangle \\
&-\frac{\kappa}{2}  \langle \hat{a}^{\dagger}  \hat{\sigma}_\mathrm{1}^{\mathrm{35}}  \rangle \\
\frac{d}{d t} \langle  \hat{\sigma}_\mathrm{1}^{\mathrm{53}} \hat{\sigma}_\mathrm{2}^{\mathrm{35}}  \rangle = & i g_{\mathrm{35}}  \langle \hat{a} \hat{\sigma}_\mathrm{1}^{\mathrm{53}}  \rangle\left(\langle  \hat{\sigma}_\mathrm{2}^{\mathrm{55}}  \rangle -\langle  \hat{\sigma}_\mathrm{2}^{\mathrm{33}}  \rangle \right) \\
&+i g_{\mathrm{35}} \langle \hat{a}^{\dagger} \hat{\sigma}_\mathrm{1}^{\mathrm{35}} \rangle \left(\langle  \hat{\sigma}_\mathrm{2}^{\mathrm{33}}  \rangle -\langle  \hat{\sigma}_\mathrm{2}^{\mathrm{55}}  \rangle \right)  -k_{\mathrm{31}} \langle  \hat{\sigma}_\mathrm{1}^{\mathrm{53}} \hat{\sigma}_\mathrm{2}^{\mathrm{35}}  \rangle \\
& -k_{\mathrm{51}} \langle  \hat{\sigma}_\mathrm{1}^{\mathrm{53}}   \hat{\sigma}_\mathrm{2}^{\mathrm{35}} \rangle  -k_{\mathrm{34}} \langle \hat{\sigma}_\mathrm{1}^{\mathrm{53}} \hat{\sigma}_\mathrm{2}^{\mathrm{35}}  \rangle -k_{\mathrm{35}} \langle  \hat{\sigma}_\mathrm{1}^{\mathrm{53}} \hat{\sigma}_\mathrm{2}^{\mathrm{35}}  \rangle \\
& -k_{\mathrm{53}} \langle \hat{\sigma}_\mathrm{1}^{\mathrm{53}} \hat{\sigma}_\mathrm{2}^{\mathrm{35}}  \rangle -k_{\mathrm{54}} \langle  \hat{\sigma}_\mathrm{1}^{\mathrm{53}} \hat{\sigma}_\mathrm{2}^{\mathrm{35}}  \rangle -\frac{1}{2} \chi_{\mathrm{34}} \langle  \hat{\sigma}_\mathrm{1}^{\mathrm{53}} \hat{\sigma}_\mathrm{2}^{\mathrm{35}}  \rangle \\
&  -2 \chi_{\mathrm{35}} \langle \hat{\sigma}_\mathrm{1}^{\mathrm{53}} \hat{\sigma}_\mathrm{2}^{\mathrm{35}}  \rangle -\frac{1}{2} \chi_{\mathrm{45}} \langle  \hat{\sigma}_\mathrm{1}^{\mathrm{53}} \hat{\sigma}_\mathrm{2}^{\mathrm{35}}  \rangle .
\end{aligned}
\end{equation*}
The contribution of thermal equilibrium photons is neglected in the above equation, as their number is negligible relative to the number of generated microwave photons.
\begin{figure}[htb]
	\includegraphics[width=0.46\textwidth]{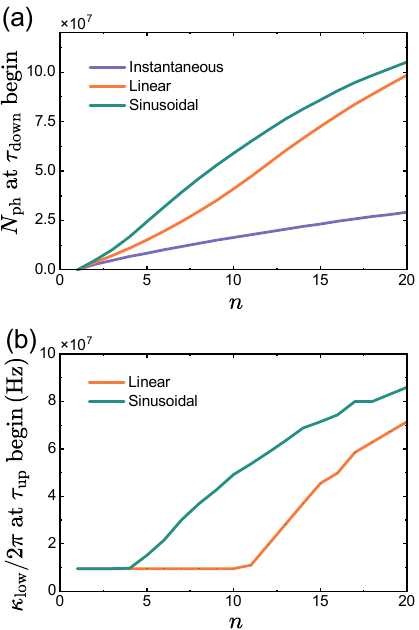}
	\caption{\label{figS1} (a) $N_{\mathrm{ph}}$ at the start of  the $\tau_{\mathrm{down}}$ (or $\tau_2$) stage in the $n$th cycle. (b) $\kappa$ at the start of the $\tau_{\mathrm{up}}$ state ($N_{\mathrm{ph}}=10^{10}$). The parameters are identical to those presented in Fig.$1$ of the main text, with $\tau_1=1 \times 10^{-8}$ s, $\tau_{\mathrm{down}}=2/ \kappa_{\mathrm{low}}$, and $\tau_{\mathrm{up}}=\tau_{\mathrm{down}}$.  The laser power is set at $2 \times 10^4$ W, with all other parameters summarized in Table.\ref{table1}.}
\end{figure}
\begin{figure}[htb]
	\includegraphics[width=0.48\textwidth]{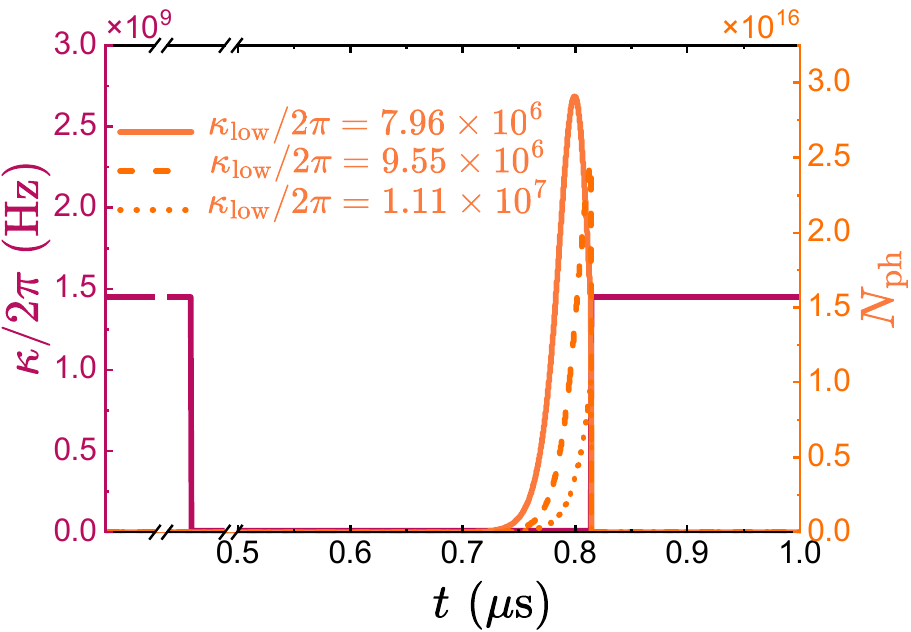}
	\caption{\label{figS2} The dissipation rate $\kappa$ and photon number $N_{\mathrm{ph}}$ vary with $t$ at different $\kappa_{\mathrm{low}}$ values under the instantaneous scheme.}
\end{figure}
\begin{figure}[htb]
	\includegraphics[width=0.47\textwidth]{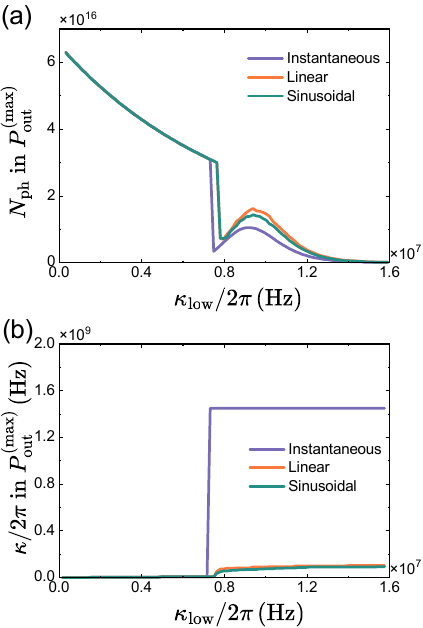}
	\caption{\label{figS3} (a) $N_{\mathrm{ph}}$ and (b) $\kappa$ at maximum power $P_{\mathrm{out}}^{(\mathrm{max})}$ for different $\kappa_{\mathrm{low}}$.   All parameters are identical to those in Fig.\hyperref[fig4]{4}.}
\end{figure}
\begin{figure*}[htb]
	\includegraphics[width=0.98\textwidth]{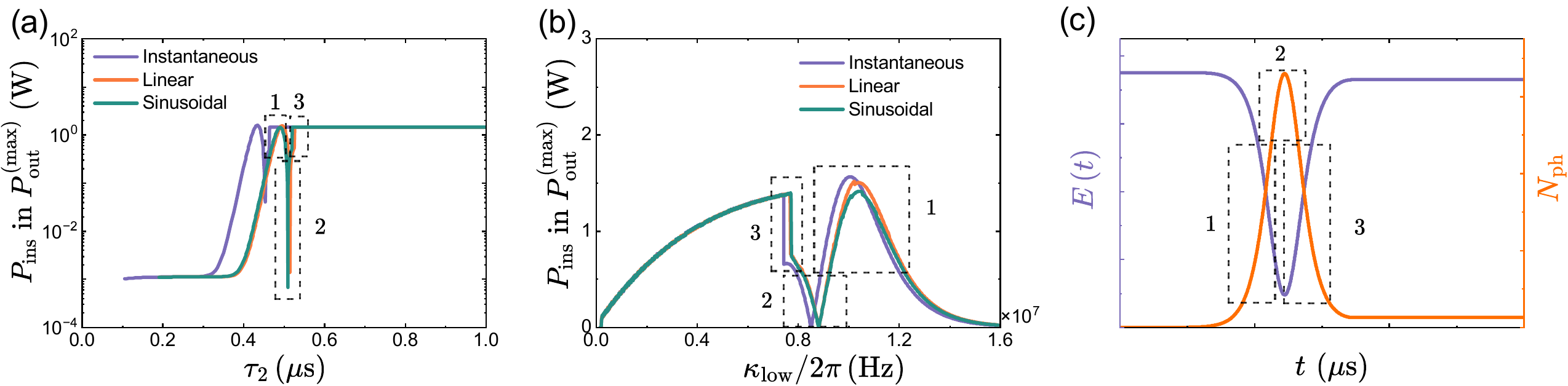}
	\caption{\label{figS4} $P_{\mathrm{ins}}$ at maximum power $P_{\mathrm{out}}^{(\mathrm{max})}$ for different (a) $\tau_{\mathrm{2}}$ or (b) $\kappa_{\mathrm{low}}$. All  parameters are identical to those in Fig.\hyperref[fig5]{5}. (c) Schematic diagram of the energy and photon number dynamics during quantum battery operation. Different regions indicate the position of photons on the pulse when the output power reaches its maximum value.}
\end{figure*}
\begingroup
\squeezetable
\begin{table*}[htb]
\begin{ruledtabular}
\begin{tabular}{cccccc}
Descriptions&Symbols&Values&Descriptions&Symbols&Values\\
\colrule
Resonator frequency & $\omega_{\mathrm{m}}/2\pi$ &  1.45 GHz &  Thermal equilibrium photon number &    $n_\mathrm{m}^\mathrm{th}$ & 4000\\
Spin-resonator coupling & $g_{\mathrm{35}}/2\pi$ & $3.66 \times 10^{-2}$ Hz\cite{PhysRevApplied.14.064017}         & Number of spins   & $N_{\mathrm{pen}}$   &  $1 \times 10^{17}$ \\
Spin transition frequency & $\omega_{\mathrm{35}}/2\pi$ &  1.45 GHz & Spin dephasing rates& $(\chi_{\mathrm{35}} \approx \chi_{\mathrm{34}} \approx \chi_{\mathrm{45}})/2\pi$  & $0.18$ MHz\cite{PhysRevApplied.14.064017}\\
Spontaneous emission rate&  $k_{\mathrm{sp}}$ & $42$ MHz\cite{10.1063/1.1499124} &Optical pumping rate&$\xi$& $3.1$ kHz/W\cite{PhysRevLett.127.053604} \\
Spin-lattice relaxation rates&$k_{\mathrm{35}}$&  $1.1 \times 10^4$ Hz\cite{PhysRevApplied.14.064017}  &  & $k_{\mathrm{34}}$ &$2.8 \times 10^4$ Hz\cite{PhysRevApplied.14.064017} \\
& $k_{\mathrm{45}}$ & $0.4 \times 10^4 $ Hz\cite{PhysRevApplied.14.064017} &&&\\
Intersystem crossing rates& $k_{\mathrm{25}}$ & $52.4$ MHz\cite{PhysRevLett.127.053604}  &  & $k_{\mathrm{24}}$ & $11$ MHz\cite{PhysRevLett.127.053604}\\
 & $k_{\mathrm{23}}$ & $5.52$ MHz\cite{PhysRevLett.127.053604} & & $k_{\mathrm{51}}$ & $2.2 \times 10^{4}$ Hz\cite{PhysRevApplied.14.064017}\\
 & $k_{\mathrm{41}}$ & $1.4 \times 10^4$ Hz\cite{PhysRevApplied.14.064017} & & $k_{\mathrm{31}}$ & $0.2 \times 10^4$ Hz\cite{PhysRevApplied.14.064017} \\
 Highest photon dissipation rate & $\kappa_{\mathrm{high}}/2\pi$    &     $1.45 \times 10^3 $ MHz & Lowest photon dissipation rate & $\kappa_{\mathrm{low}}/2\pi$    &     $9.55   $ MHz
\end{tabular}
\end{ruledtabular}
\caption{\label{table1}Summary of the parameters involved in the dynamics of the pentacene molecular quantum battery in Fig.\hyperref[figS1]{S1}.}
\end{table*}
\endgroup

\section{SIMULATION OF THE DISCHARGING DYNAMICS}

In the three regulation schemes, the photon number $N_{\mathrm{ph}}$ at the beginning of each control cycle increases progressively, as shown in Fig.\hyperref[figS1]{S1(a)}. This behavior accounts for the observed reduction in the FWHM in Fig.2 of the main text. In Fig.\hyperref[figS1]{S1(b)}, for the linear scheme, $\kappa/2\pi = \kappa_{\mathrm{low}}/2\pi$ ($9.55 \times 10^6$ Hz)  during the first $10$ cycles upon entering stage $\tau_{up}$, but $\kappa > \kappa_{\mathrm{low}}$ thereafter. This indicates that the photon number reached the predefined threshold $10^{10}$ in stage $\tau_{\mathrm{down}}$ after $10$ cycles, bypassing stage $\tau_{\mathrm{low}}$ and without completing stage $\tau_{\mathrm{down}}$ ($\tau_{\mathrm{up}}$). Similarly, for the sinusoidal scheme, the photon number attained the threshold  in stage $\tau_{\mathrm{down}}$ after the $4$th cycle. Since stage $\tau_{\mathrm{up}}$ was not fully completed in either case, the maximum photon number began to decrease after the $10$th cycle for the linear scheme and after the $4$th cycle for the sinusoidal scheme, as illustrated in Fig.\hyperref[fig2]{2(e)} and Fig.\hyperref[fig2]{2(f)} in the main text.

Fig.\hyperref[figS2]{S2} presents the evolution of both the dissipation rate and photon population over time, under the transient protocol with different $\kappa_{\mathrm{low}}$. When $\kappa/2\pi=7.96\times 10^6$ Hz, the microwave photon pulse can still reach its peak; when $\kappa/2\pi=9.55\times 10^6$ Hz, the number of generated microwave photons cannot reach the peak of the pulse; and when $\kappa/2\pi=1.11\times 10^7$ Hz, the number of generated microwave photons is less than half of the peak. Therefore, near $\kappa/2\pi=9.55\times 10^6$ Hz, especially after $\kappa/2\pi>9.55\times 10^6$ Hz, the microwave optical pulses become increasingly incomplete, resulting in the rapid decline of the maximum photon number shown in Fig.\hyperref[fig4]{4(e)} of the main text.

Fig.\hyperref[figS3]{S3} illustrates the photon number $N_{\mathrm{ph}}$ and dissipation rate $\kappa$ at maximum power across various values of parameter $\kappa_{\mathrm{low}}$. When $\kappa/2\pi < 7.16 \times 10^6$ Hz, the first pulse can be fully observed, allowing $N_{\mathrm{ph}}$ to reach its maximum value, with $\kappa = \kappa_{\mathrm{low}}$ at $P_{\mathrm{out}}^{(\mathrm{max})}$. Subsequently, only a partial area of the first pulse can be obtained gradually, leading to a rapid decrease in $N_{\mathrm{ph}}$ while $\kappa$ at $P_{\mathrm{out}}^{(\mathrm{max})}$ increases accordingly. The instantaneous scheme corresponds directly to $\kappa = \kappa_{\mathrm{high}}$, and consequently, as illustrated in Fig.\hyperref[fig4]{4(f)}, the output power achieves its maximum value at $\kappa/2\pi=9.55 \times 10^6$ Hz.

In Fig.\hyperref[figS4]{S4(a)} and Fig.\hyperref[figS4]{S4(b)}, as parameters $\tau_{\mathrm{2}}$ or $\kappa_{\mathrm{low}}$ increase, when the position of the maximum output power $P_{\mathrm{out}}^{(\mathrm{max})}$ approaches the peak of the microwave photon pulse, the corresponding instantaneous power $P_{\mathrm{ins}}$ first decreases and then increases. The underlying mechanism of this phenomenon can be understood with reference to Fig.\hyperref[figS4]{S4(c)}. Instantaneous power is the rate of change of energy, and the change in energy is basically opposite to the change in the number of photons. 

Taking $\kappa_{\mathrm{low}}$ as an example, as $\kappa_{\mathrm{low}}$ gradually increases, the acquired microwave pulse becomes increasingly incomplete. The position of $P_{\mathrm{out}}^{(\mathrm{max})}$ shifts from the vicinity of the peak of the microwave optical pulse to the transition point between stage $\tau_{\mathrm{low}}$ and stage $\tau_{\mathrm{up}}$ (or new stage $\tau_{\mathrm{1}}$)—corresponding to the boundary between region $2$ and region $3$ in Fig.\hyperref[figS4]{S4(c)}. This shift aligns with a sudden change in instantaneous power observed in region $3$ of Fig.\hyperref[figS4]{S4(b)}. As $\kappa_{\mathrm{low}}$ continues to increase, the position of $P_{\mathrm{out}}^{(\mathrm{max})}$ shifts through the peak of the microwave optical pulse, resulting in an initial decrease followed by an increase in instantaneous power (see region $2$ in Fig.\hyperref[figS4]{S4(b)} and Fig.\hyperref[figS4]{S4(c)}). When the photon number at $P_{\mathrm{out}}^{(\mathrm{max})}$ reaches approximately half of the peak value of the entire microwave optical pulse, the instantaneous power exhibits a transient increase followed by a decrease (see region $1$ in Fig.\hyperref[figS4]{S4(b)} and Fig.\hyperref[figS4]{S4(c)}).

As the parameter $\tau_{\mathrm{2}}$ increases, the sequence in which each region appears is reversed. In the process of progressively generating a complete microwave optical pulse, region $1$ is encountered first, followed by regions $2$ and $3$, as illustrated in Fig.\hyperref[figS4]{S4(a)}. The variation in instantaneous power constitutes the primary factor responsible for the observed changes in parameters $\eta_{\mathrm{work}}$ and $\eta_{\mathrm{power}}$ depicted in Fig.\hyperref[fig5]{5(c)} and \hyperref[fig5]{5(d)}.

\bibliography{Ref}



\end{document}